\documentclass[conference]{IEEEtran}
\IEEEoverridecommandlockouts
\let\oldbibliography\thebibliography
\renewcommand{\thebibliography}[1]{%
  \oldbibliography{#1}%
  \setlength{\itemsep}{0pt}}   

\usepackage{tikz}
\usetikzlibrary{shapes.geometric, arrows}
\usepackage{cite}
\usepackage{amsmath,amssymb,amsfonts}
\usepackage{algpseudocode}
\usepackage{booktabs}
\usepackage{graphicx}
\usepackage{textcomp}
\usepackage{xcolor}
\usepackage{url}
\usepackage{wrapfig}
\usepackage{balance}
\usepackage{subcaption}
\usepackage[font=footnotesize,labelfont=bf]{caption}
\usepackage{booktabs,tabularx,graphicx}

\begin{document}
\title{
Predicting Cascading Failures in Power Systems using Machine Learning }

\author{

{\tt\small}
Samita Rani Pani\\
KIIT Deemed to be University \\
Bhubaneswar, Odissa, India \\
{\tt\small samitapani.fel@kiit.ac.in\vspace*{-0.1 cm}}

\and
{\tt\small}
Pallav Kumar Bera\\
Western Kentucky University\\
Bowling Green, KY, USA \\
{\tt\small pallav.bera@wku.edu\vspace*{-0.1 cm}}

\and
{\tt\small}
Rajat Kanti Samal\\
VSSUT Burla \\
 Odissa, India \\
{\tt\small rksamal\_ee@vssut.ac.in\vspace*{-0.1 cm}}

\thanks{© 2024 IEEE. Personal use of this material is permitted. Permission from IEEE must be obtained for all other uses, in any current or future media, including reprinting/republishing this material for advertising or promotional purposes, creating new collective works, for resale or redistribution to servers or lists, or reuse of any copyrighted component of this work in other works.
}}

\maketitle
\begin{abstract}

Cascading failure studies help assess and enhance the robustness of power systems against severe power outages. Onset time is a critical parameter in the analysis and management of power system stability and reliability, representing the timeframe within which initial disturbances may lead to subsequent cascading failures. In this paper, different traditional machine learning algorithms are used to predict the onset time of cascading failures. The prediction task is articulated as a multi-class classification problem, employing machine learning algorithms. The results on the UIUC 150-Bus power system data available publicly show high classification accuracy with Random Forest. The hyperparameters of the Random Forest classifier are tuned using Bayesian Optimization. This study highlights the potential of machine learning models in predicting cascading failures, providing a foundation for the development of more resilient power systems.
\end{abstract}

\begin{IEEEkeywords}
Cascading failures, Onset time, Decision Tree, Support Vector Machines, Random Forest, Neural Network, Naive Bayes, Logistic Regression, k-Nearest Neighbor
\vspace{-0.1cm}
\end{IEEEkeywords}

\section{Introduction}
The increasing complexity and interconnectivity of modern power grids have made them more vulnerable to cascading failures, where an initial disturbance can lead to widespread outages. Over the past three decades, significant cascading failures in power systems have underscored grid vulnerabilities and the need for robust planning. For instance, in 1996, Western North America experienced two blackouts due to high demand and equipment failures, affecting millions in the U.S. and Canada \cite{western1996}. Similarly, the 2003 Northeast Blackout impacted approximately 50 million people in the U.S. and Ontario due to a software bug and operator errors\cite{us-canada2004}. More recently, Pakistan's 2021 blackout, caused by an engineering fault in a power plant, affected 90\% of the population \cite{masood2021}. Additionally, the 2012 India Blackout, the largest in history, affected over 620 million due to states overdrawing power \cite{india2012}.  On March 11, 1999, Southern Brazil experienced a massive blackout affecting approximately 75 million people due to a lightning strike that caused a cascading failure in the electrical grid \cite{brazil}. These events highlight the complexity of modern power grids and the necessity for continuous investment in infrastructure and emergency response strategies.

 Previous studies have explored various methods to assess and improve power system reliability and vulnerability. For instance, reliability assessment techniques for distribution power networks have been developed to enhance grid robustness and prevent outages \cite{samitaodicon}. Further, graph-theoretic approaches have been employed to evaluate power grid vulnerabilities to transmission line outages, offering insights into potential weak points in the grid \cite{samitaICICCSP}. Assessing power system vulnerabilities under N-1 contingency conditions has been a focal point in identifying and mitigating risks associated with single component failures \cite{samita_vulnerability}. Evaluating power grid vulnerability indices that account for wind power uncertainty also provides a comprehensive understanding of grid stability in the face of renewable energy integration \cite{samita_segan}. Researchers have proposed the use of machine learning techniques in various areas of power systems to enhance their efficiency and reliability \cite{systempallav}. Predictive analysis of cascades in power systems is crucial for mitigating risks of cascading failures and blackouts. This may involve the prediction of cascade size, prediction of cascade trajectory and regions at risk, and prediction of cascade evolution and its temporal characteristics \cite{review}. 

First, cascade size prediction is an essential aspect of understanding and mitigating cascading failures in power systems. This process involves estimating the potential magnitude of a cascade once the risk has been identified. The prediction is typically approached by analyzing the distribution of cascade sizes based on factors such as the number of component failures, the quantity of load dropped, or the total number of clients impacted.
Second, predicting cascade paths and areas at risk involves identifying the specific components or regions within a power system that are likely to fail following an initial disturbance. This is challenging due to the complex interactions between system components. Research studies focus on predicting the next components to fail, the sequence of multiple failures, and the entire path of the cascade. 
Third, the prediction of cascade evolution and its temporal characteristics focuses on understanding the phases of cascading failures over time. This involves categorizing the propagation of failures into distinct phases and estimating the time available for operators to respond effectively. Key methods include data-driven models like Markov models and neural networks, which classify the urgency of situations and predict the onset time of critical phases. Despite its importance, this area has limited research, indicating a need for further exploration to improve the comprehension and management of cascading failures \cite{fang}.

Reference \cite{markov} study proposes a phase transition model for cascading network failure, emphasizing the critical points at which the system shifts from stability to cascading failures. This model is instrumental in understanding the dynamic behavior of power systems under stress and identifying the phases of failure propagation.
Reference \cite{graph} introduces the use of graph neural networks (GNNs) for predicting power failure cascades. GNNs effectively capture the complex dependencies and topologies in power grids, enhancing the accuracy of predicting cascading failure paths and the affected areas.
Reference \cite{fang} utilizes a classifier based on neural networks to forecast when power systems cascading failures would occur. The classifier helps in estimating the critical phases and provides operators with crucial time to react and mitigate the impacts of these failures. The large-area cascading failure prevention indication for power systems is the onset time, which is the point at which the cascading failure's propagation rate starts to climb quickly.

This study focuses on predicting the onset time of cascading failures using various traditional machine learning algorithms. The onset time is a critical parameter representing the timeframe within which initial disturbances can lead to subsequent cascading failures. By accurately predicting this onset time, we aim to better the stability and reliability of power systems.
The following is a summary of the work's highlights: 

{\raisebox{-0.4\height}{\scalebox{1.6}{\textbullet}}} The study applies various traditional machine learning algorithms to predict the onset time of cascading failures in power systems, highlighting their comparative performance and suitability.

{\raisebox{-0.4\height}{\scalebox{1.6}{\textbullet}}} Optimization of the machine learning model’s hyperparameters using Bayesian Optimization, ensuring the best possible predictive performance.

{\raisebox{-0.4\height}{\scalebox{1.6}{\textbullet}}} Proposes a robust classifier capable of predicting failure onset time based on input power matrix data, aimed at improving power system stability and reliability.

The remaining paper is organized as follows: Section II describes the onset time of cascading failure, discusses the Random Forest model applied to predict the onset of cascading failures, and the hyperparameter optimization technique employed. Section III presents the results of the experiments, including a comparative analysis of model performances. Section IV concludes the paper, summarizing the findings.

\section{Methodology}

This section introduces the concept of onset time, as defined in literature and formulates the onset time prediction as a three-class classification situation.  A machine learning-based classifier that predicts failure onset time depending on the input power matrix is proposed. The hyperparameters of the machine learning algorithm are tuned using Bayesian Optimization.

\subsection{Data Description and Classification Criteria}
The dataset provided in \cite{fang} has been utilized in this work. Two separate power lines are designated as malfunctioning and must be eliminated in order to build the dataset by applying the notion of N-2 (k=2) security. For the purpose of training and testing the machine learning classifiers, thus 20503 failure propagation profiles are obtained.  It includes 3704 non-critical events, 14700 critical events, and 2099 relative critical events. By using the approach outlined in reference \cite{liu}, the onset time is determined. The onset time $t_{c1}$ and $t_{c2}$ are set to 100 and 1000 minutes, respectively, in order to establish 3 time intervals for the 3 onset time classes.

\begin{figure}[htbp]
\centerline{\includegraphics[width=2.5 in, height= 1.2 in]{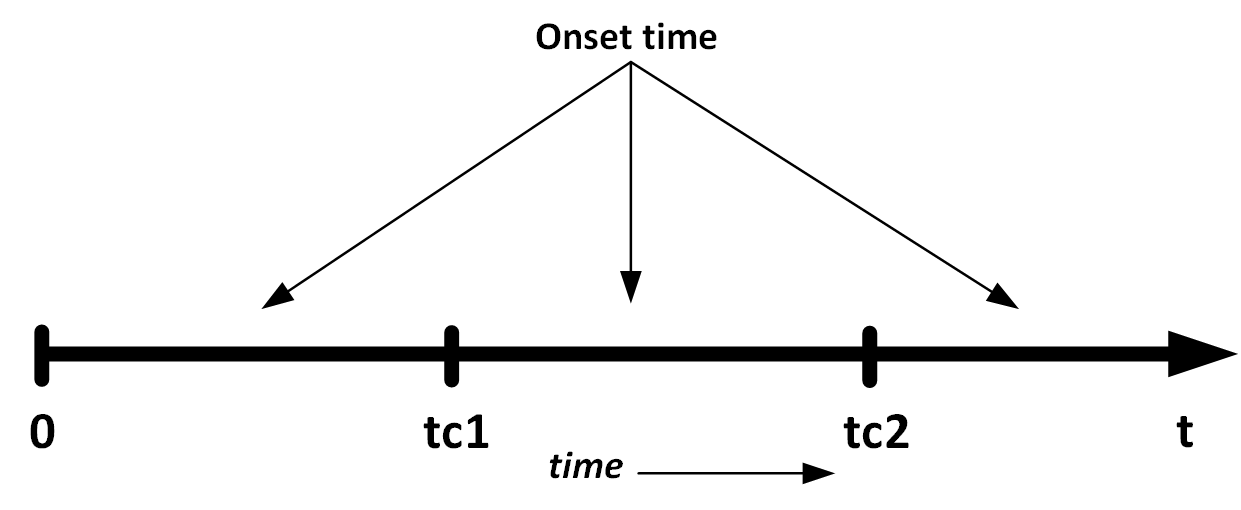}}\caption{Three time intervals: $[0, t_{c1})$, $[t_{c1}, t_{c2})$, and $[t_{c2}, +\infty)$ representing  critical, relatively critical, and non-critical cases.}
\label{onset time}
\end{figure}

Instead of directly determining the onset time value, the aim is to develop a machine learning algorithm-based classifier to categorize the onset time. Based on the duration, the onset time is classified into 3 categories: critical, relatively critical, and non-critical. These categories indicate the urgency of protective actions needed to stop a rapid and significant failure propagation in a power system. To distinguish these categories, the two critical time points, $t_{c1}$ and $t_{c2}$ are defined, resulting in the 3 intervals: $[0, t_{c1})$, $[t_{c1}, t_{c2})$, and $[t_{c2}, +\infty)$ (figure \ref{onset time}). The goal of the machine learning classifier is to predict the interval in which the onset time falls. If no onset time is detected in the failure propagation profile, it is classified as a non-critical case. The input variable, or the predictor for the machine learning classifier, is derived from the power matrix difference $\Delta M_p$. This matrix is derived from the initial steady state and after the removal of a certain number of power lines (k). The power matrix difference $\Delta M_p$ is calculated to capture the initial state differences. $\Delta M_p$(i, j) is zero if there's no transmission line between buses i and j.

\subsection{Random Forest}
The onset times are used to train various classifiers, including Neural Network, Decision Tree, Random Forest, Naive Bayes, Support Vector Machines, Logistic Regression, and k-Nearest Neighbors.

Random Forest is an ensemble learning-based classification technique that uses random feature selection and bootstrapping (choosing portions of the data at random) to generate numerous decision trees from subsets of data. A majority vote among all the trees determines the final prediction, with each tree forecasting a class label \cite{breiman2001random} (figure 2). Because of this method's increased accuracy and decreased chance of overfitting, Random Forest is a reliable and well-liked classifier. For a given input sample \( p \), the predicted class is expressed as follows:

\[
\hat{q} = \text{argmax}_i \sum_{j=1}^{N_T} I\left( T_j(p) = i \right)
\]

where \( \hat{q} \) is the anticipated class label for the incoming sample \(p \), \( N_T \) is the total number of decision trees, and \( T_j(p) \) is the predicted class label by the \( j \)-th decision tree. The indicator function \( I \) returns 1 if the condition in parentheses is true and 0 otherwise.

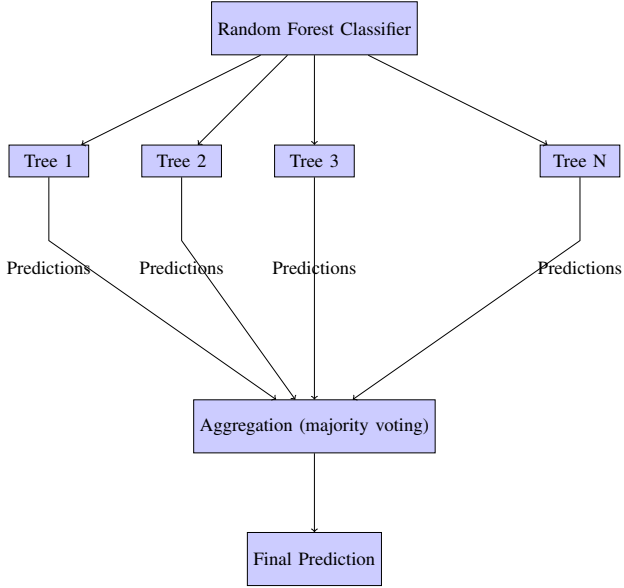
\begin{figure}[ht]
    \centering
\resizebox{9.0cm}{!}{
\begin{tikzpicture}[node distance=2.5cm, every node/.style={draw, fill=blue!20, rectangle, minimum height=1cm, minimum width=2.5cm}]
    \tikzstyle{main} = [draw, fill=blue!20, rectangle, minimum height=1cm, minimum width=2.5cm]

    \tikzstyle{tree} = [draw, fill=blue!20, rectangle, minimum height=0.6cm, minimum width=1.5cm]

    \node[main] (rf) {Random Forest Classifier};
    \node[tree] (t1) [below of=rf, xshift=-5cm] {Tree 1};
    \node[tree] (t2) [below of=rf, xshift=-2.5cm] {Tree 2};
    \node[tree] (t3) [below of=rf, xshift=0cm] {Tree 3};
    \node[tree] (tn) [below of=rf, xshift=5cm] {Tree N};
    \node[main] (agg) [below of=t3, yshift=-2.5cm] {Aggregation (majority voting)};
    \node[main] (fp) [below of=agg] {Final Prediction};

    \draw[->] (rf) -- (t1);
    \draw[->] (rf) -- (t2);
    \draw[->] (rf) -- (t3);
    \draw[->] (rf) -- (tn);
    \draw[->] (t1) -- +(0,-1.5) node[anchor=north,draw=none,fill=none] {Predictions} -- (agg);
    \draw[->] (t2) -- +(0,-1.5) node[anchor=north,draw=none,fill=none] {Predictions} -- (agg);
    \draw[->] (t3) -- +(0,-1.5) node[anchor=north,draw=none,fill=none] {Predictions} -- (agg);
    \draw[->] (tn) -- +(0,-1.5) node[anchor=north,draw=none,fill=none] {Predictions} -- (agg);
    \draw[->] (agg) -- (fp);
\end{tikzpicture}
  }
    \caption{General Random Forest Classifier Diagram}
\end{figure}
\begin{table*}[h!]
\centering
\caption{Performance comparison of N-2 with different models}
\begin{tabular}{lcccc}
\toprule
\textbf{Model} & \textbf{ Training Accuracy} & \textbf{ Test Accuracy} & \textbf{Training Time (s)} & \textbf{Testing Time (s)} \\
\midrule
Decision Tree & 1.0 & 0.8766 & 30.1860 & 0.8789 \\
Random Forest & 1.0 & 0.9010 & 198.0825 & 2.0514 \\
Naive Bayes & 0.7522 & 0.7501 & 3.8943 & 2.9879 \\
k-Nearest Neighbor & 0.8042 & 0.7848 & 1.1150 & 17.1373 \\
Logistic Regression & 0.8244 & 0.8132 & 12.6935 & 2.3645\\
Neural Network & 0.9488 & 0.8997 & 1408.41 & 0.0200 \\
\bottomrule
\end{tabular}
\label{table1}
\vspace{5mm}
\end{table*}

\begin{table}[h!]
    \centering
    \caption{Hyperparameter Grid used and Bayesian Optimization Results obtained for Random Forest Classifier}
    \label{hyper}
    \begin{tabular}{>{\bfseries}l l l}
        \toprule
        \textbf{Hyperparameter} & \textbf{Values} & \textbf{Best Value} \\
        \midrule
        no. of trees          & 200,400,600,800,1000 & 600 \\
        max. depth of tree            & 10, 20, 30, 40 & 40 \\
        min. samples required to split    & 2, 5, 10, 15 & 10 \\
        min. samples in new leaves     & 1, 2, 4, 6, 8 & 1 \\
        no. of features          & sqrt, log2 & sqrt \\
        bootstrap              & True, False & False \\
        \midrule
        \textbf{Test Accuracy} & \multicolumn{2}{c}{0.9060} \\
        \bottomrule
    \end{tabular}
\end{table}

\begin{table*}[h]
\centering
\caption{Performance comparison of different models}
\begin{tabular}{lcccccc}
\toprule
\textbf{Model} & \textbf{N-2 Modified Accuracy} & \textbf{N-3 Accuracy} & \textbf{N-4 Accuracy} & \textbf{N-5 Accuracy} & \textbf{N-6 Accuracy} \\
\midrule
Decision Tree & 0.6686 & 0.8508 & 0.8356 & 0.8182 & 0.7886 \\
Random Forest & 0.6816 & 0.9220 & 0.9490 & 0.9672 & 0.9668 \\
Naive Bayes & 0.5458 & 0.7738 & 0.8548 & 0.9146 & 0.9380 \\
k-Nearest Neighbor & 0.5992 & 0.7584 & 0.8232 & 0.8724 & 0.9034 \\
Logistic Regression & 0.6976 & 0.8418 & 0.8612 & 0.8854 & 0.9008\\
Neural Network & 0.6474 & 0.8826 & 0.9170 & 0.9418 & 0.9508 \\
\bottomrule
\end{tabular}
\label{table2}
\vspace{5mm}
\end{table*}

\subsection{Hyperparameter Optimization}
Hyperparameter optimization involves tuning the settings that control the behavior of a machine learning algorithm to improve model performance or accuracy. Search methods like grid search, random search, and Bayesian optimization are often used to determine the best hyperparameters, thereby enhancing the machine learning model's ability to generalize from training data.

Bayesian hyperparameter optimization uses Bayesian statistics to efficiently search for optimal hyperparameters \cite{pallav_access}. By building a probabilistic model of the objective function, it predicts the performance at different settings and guides the search toward the most promising areas, reducing the number of evaluations needed.  The Bayesian Optimization algorithm is described below.

\begin{enumerate}
    \item \textbf{Input:} Objective function $f(\mathbf{x})$, initial data $\mathcal{D}$, max iterations $N$, acquisition function $\alpha(\mathbf{x}|\mathcal{D})$
    \item \textbf{For} $n = n_0+1$ to $N$:
    \begin{enumerate}
        \item Fit probabilistic model to $\mathcal{D}$
        \item $\mathbf{x}_n = \arg\max_{\mathbf{x}} \alpha(\mathbf{x}|\mathcal{D})$
        \item Evaluate $f(\mathbf{x}_n)$ and update $\mathcal{D}$
    \end{enumerate}
    \item \textbf{Output:} $\mathbf{x}^* = \arg\min_{\mathbf{x}_i \in \mathcal{D}} f(\mathbf{x}_i)$
\end{enumerate}

\begin{figure}[h]
    \centering
    \includegraphics[width=0.6\linewidth]{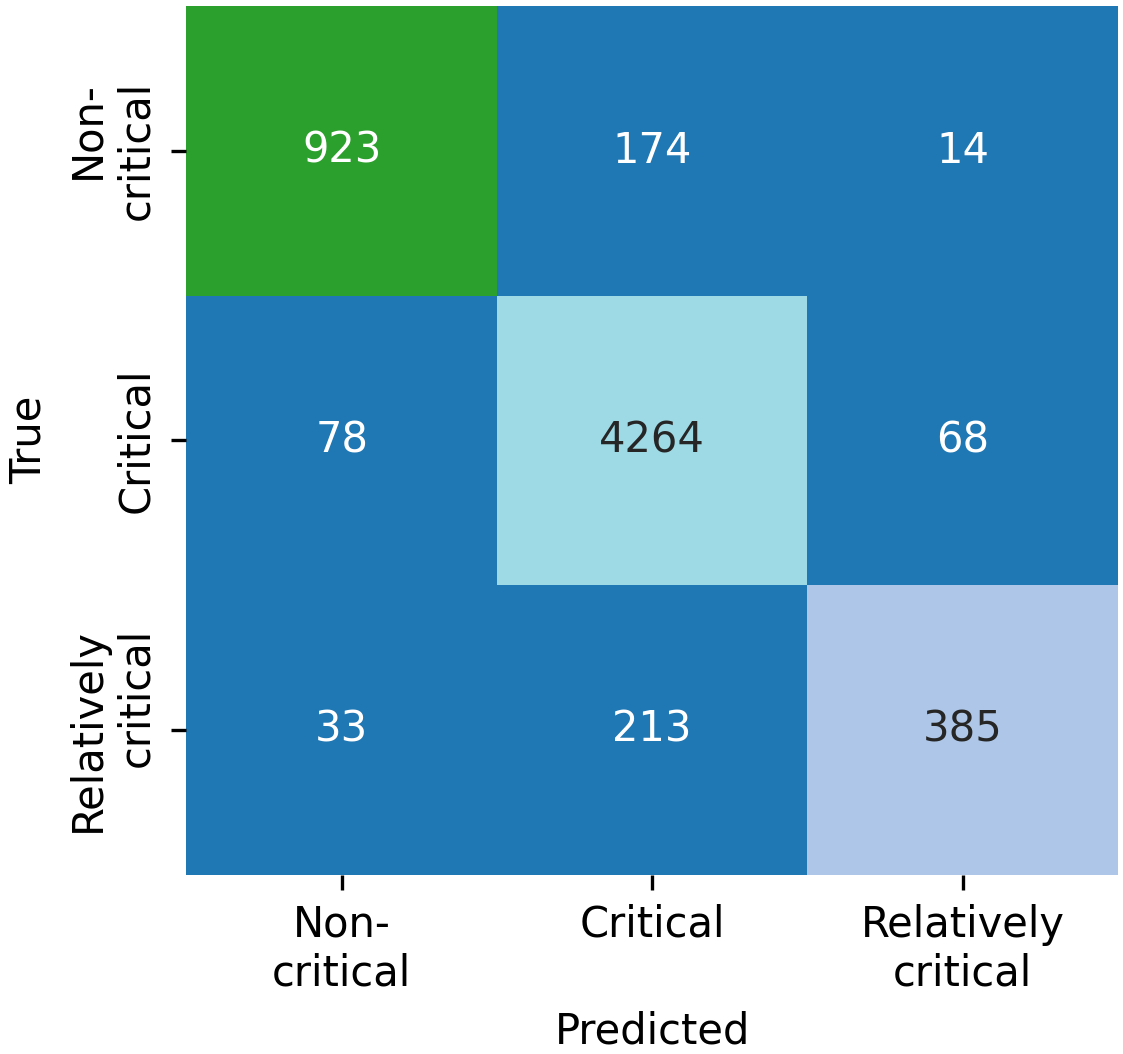} 
    \caption{Confusion Matrix for Random Forest with Optimized Parameters}
    \label{con}
\end{figure}

\section{Results}
The data split ratio is 70\% for training and 30\% for testing. The accuracy of the  classification models is calculated as:
\[
\text{Accuracy} = \frac{\text{Number of correct predictions}}{\text{Total number of predictions}}
\]

The table \ref{table1} provides a comparison of various models based on their training accuracy, test accuracy, training time, and testing time. Overall, the Random Forest model provides the best test accuracy of 90.10\% but requires more training time compared to some other models. Neural Networks also perform well but at a high computational cost during training. Naive Bayes and k-Nearest Neighbor are faster to train but do not achieve the same level of accuracy as other models.

BayesSearchCV from the skopt library is used to optimize the hyperparameters. Specifically, it performed 50 iterations (n\_iter=50) of the search, utilized 3-fold cross-validation (cv=3) to evaluate each parameter set, and leveraged parallel processing by utilizing all available processors (n\_jobs=-1) to expedite the computation.
The hyperparameters: `no. of trees', `max. depth of tree',  `min. samples required to split',  `bootstrap',  `min. samples in new leaves', and `no. of features' of the Random Forest classifier are tuned using Bayesian Optimization. The different settings of these hyperparameters and the best value obtained are shown in Table \ref{hyper}. Hyperparameter optimization improves the accuracy from 90.10\% to 90.60\%. This performance exceeds the previous best accuracy of 89.93\% reported in \cite{fang}, achieved on this dataset using a Neural Network.

The confusion matrix in figure \ref{con} provides a detailed breakdown of the classification performance of the Random Forest model used in the study. It compares the actual urgency levels of cascading failures (ground truth) with the predicted levels made by the model. The matrix helps in understanding the accuracy and types of errors made by the classifier.

\begin{figure}[h]
    \centering
    \includegraphics[width=\linewidth]{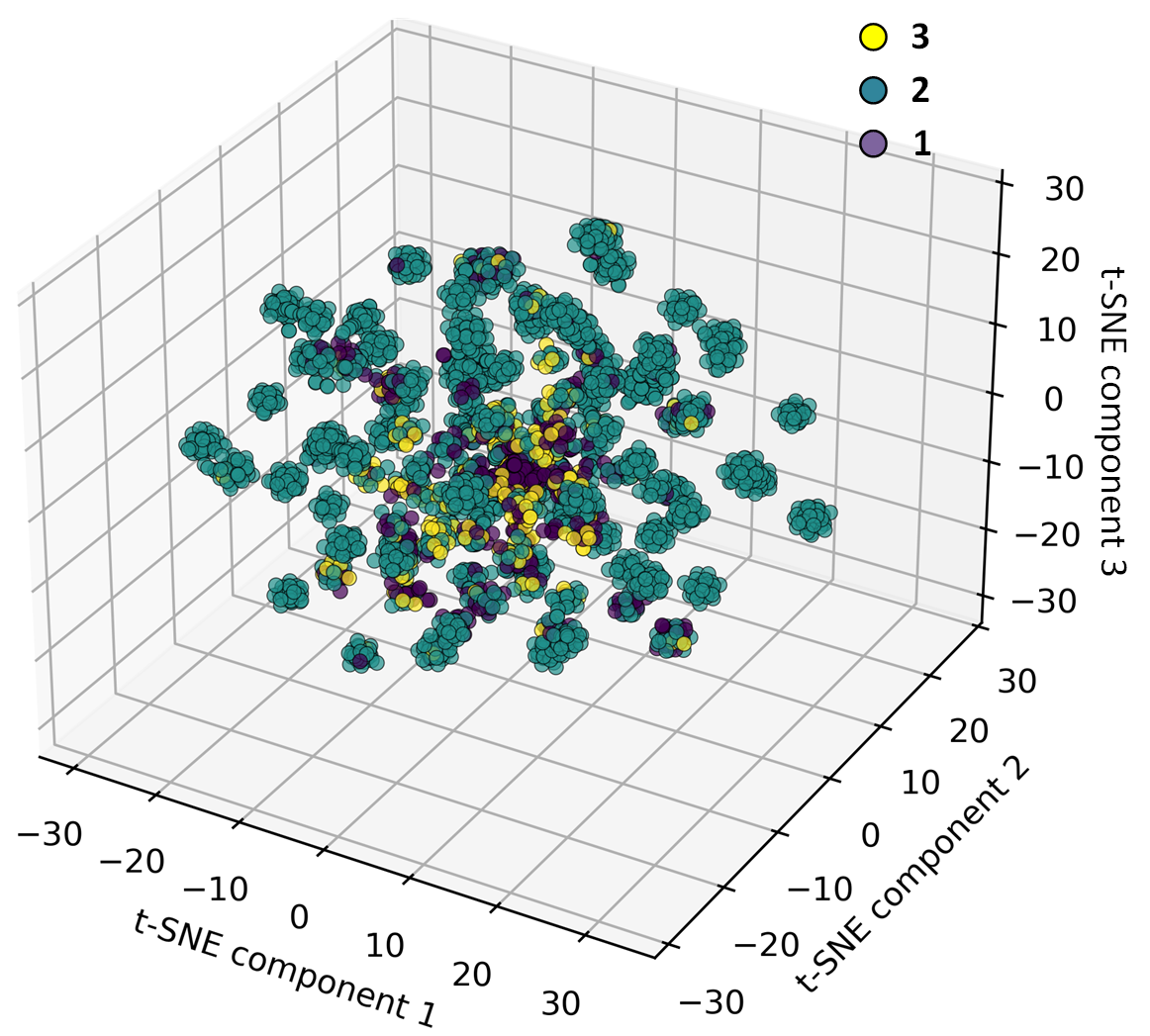}\caption{t-SNE plot of input vectors  in 3D space.  1, 2, and 3 represent non-critical, critical, and relatively critical classes.}
    \label{tsne}
\end{figure}

The t-SNE plot in figure \ref{tsne} visualizes the input vectors in a 3D space, clustering them into three distinct classes labeled 1, 2, and 3. These classes correspond to non-critical, critical, and relatively critical cases, respectively. 
This plot helps in understanding how well the input features can distinguish between different urgency levels of cascading failures in the power system. The separation between the clusters indicates that the chosen features and the machine learning model can effectively differentiate between the different levels of urgency, which is crucial for timely and appropriate responses to potential failures.

The table \ref{table2} presents a performance comparison of different machine learning models across various scenarios, specifically N-2 Modified (obtained by modifying the original system parameters), N-3 (k increased to 3), N-4, N-5, and N-6. 
Overall, the Random Forest model outperforms the other models in most cases, particularly with higher k values.  Logistic Regression and Neural Networks also demonstrate strong performance across various k values.

\section{Conclusion}
In this study, a comprehensive approach to predict the onset time of cascading failures in power systems using various traditional machine learning algorithms is presented. The onset time, categorized into critical, relatively critical, and non-critical cases, serves as a critical parameter for determining the urgency of protective actions. By leveraging the differences in power matrices before and after the removal of power lines,  a classifier capable of accurately predicting the onset time of failures is developed.

The results indicate that the Random Forest classifier, optimized using Bayesian Optimization, outperforms other models in terms of accuracy and robustness. The achieved accuracy of  90.60\% surpasses the previous record for the same dataset by 0.67\%. The findings underscore the importance of accurate onset time prediction in enhancing the stability and reliability of power systems, thereby contributing to the prevention of large-scale cascading failures. Future work may explore the integration of more advanced machine learning techniques and real-time data to further improve the predictive performance and applicability of the proposed approach.

\balance
\bibliographystyle{IEEEtran}
\bibliography{ref}
\end{document}